\documentclass[
    ,final            
  ]
  {aipproc}

\layoutstyle{6x9}

\begin{document}

\title{Sherlock: An Automated Follow-Up Telescope for Wide-Field Transit Searches}

\author{Lewis Kotredes, David Charbonneau, Dagny L. Looper \\ and Francis T. O'Donovan}{
  address={California Institute of Technology, M/C 105-24, 1200 E California Blvd, Pasadena, CA 91125; ltk,dc,ftod@astro.caltech.edu, dagny@caltech.edu}
}

\begin{abstract}
The most significant challenge currently facing photometric surveys for 
transiting gas-giant planets is that of confusion with eclipsing binary 
systems that mimic the photometric signature. A simple way to reject most 
forms of these false positives is high-precision, rapid-cadence monitoring of 
the suspected transit at higher angular resolution and in several filters. We 
are currently building a system that will perform higher-angular-resolution, 
multi-color follow-up observations of candidate systems identified by Sleuth 
(our wide-field transit survey instrument at Palomar), and its two twin system
instruments in Tenerife and northern Arizona.
\end{abstract}

\maketitle

\section{Introduction}

Wide-field photometric surveys for transits of short-period gas-giant planets 
consist of several months of single-band observations of typically 5000 
targets in a six-degree-square field of view. A number of such surveys, 
including the network consisting of Sleuth~\footnote{
\url{http://www.astro.caltech.edu/~ftod/sleuth.html}} (Palomar, PI: D. 
Charbonneau, see \cite{7}), STARE~\footnote{
\url{http://www.hao.ucar.edu/public/research/stare/stare.html}} 
(Tenerife, PI: T. Brown) and PSST (Lowell, PI: E. Dunham) have produced 
several candidates with light curves very similar to that of HD 209458 
\cite{1}. The 
most significant challenge facing such surveys is not the difficulty of 
obtaining the requisite precision and phase coverage, but rather the ability 
to rule out the large number of false positives that are typically 
encountered. These false positives can be removed through high-resolution 
spectroscopy, as was done in the case of the OGLE-III transit candidates 
\cite{2,4}. However, for relatively bright stars, there is an easier way to 
reject such candidates.

Brown \cite{3} discusses various transit-like signals typically identified by 
searches similar to Sleuth and classifies them into a number of specific 
types. There are three primary sources of these false positives. The first of 
these are binary stars undergoing a grazing eclipse (MSU). The second and 
third types involve blends between a binary star undergoing a deep eclipse 
and a third star, either a foreground object (MSDF) or a third member of the 
system (MSDT). Using a typical transit campaign for the STARE instrument, 
Brown calculates that for every 10000 stars observed, 0.39 planets will be 
detected with three transits. However, 2.27 false positives of type MSU will 
also appear, 1.26 false positives of type MSDF will be present, and 0.98 
candidates of type MSDT will be observed. Combining these numbers, Brown 
finds that a transit survey will observe more than 10 false positives for 
every true planet. This is demonstrated by Figure 1, which shows the 
probability of finding different classes of object at a given transit depth. 
In the regime of 1-2 \% eclipse depths that we study, these three sources of 
false positives are at least as plentiful as the planets we seek.

\begin{figure}
\includegraphics[width=0.65\textwidth]{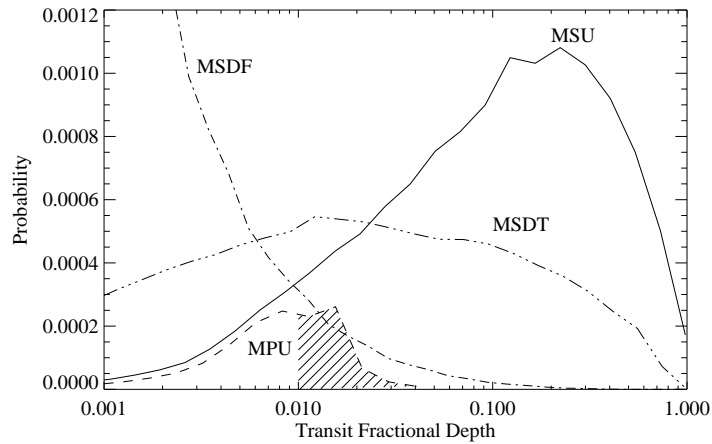}
\label{Brown}
\caption{Taken from~\cite{3} (courtesy T. Brown), the marginalized probabilities of detection of 
a transit per unit log transit depth. The shaded area is represents planets 
detectable in current surveys. Figure labels are explained in the text.}
\end{figure}

\begin{figure}
\includegraphics[width=0.65\textwidth]{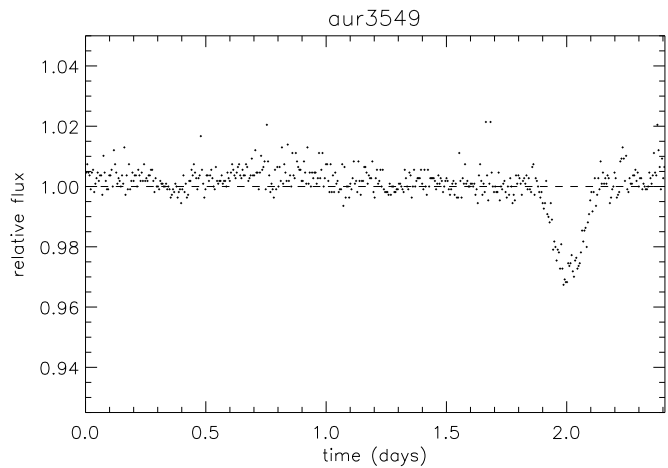}
\label{Auriga}
\caption{Phased R-band light curve of a star in Auriga as observed with the 
PSST by E. Dunham \& G. Mandushev (Lowell Observatory). Based on this light 
curve alone, this appears to be a reasonable candidate, as the period, 
duration, and depth are consistent with the passage of an inflated gas-giant 
across a Sun-like star. However, follow-up photometry shows this object to be 
a blend containing an eclipsing binary (see Figure 3).}
\end{figure}

\begin{figure}
\includegraphics[width=0.65\textwidth]{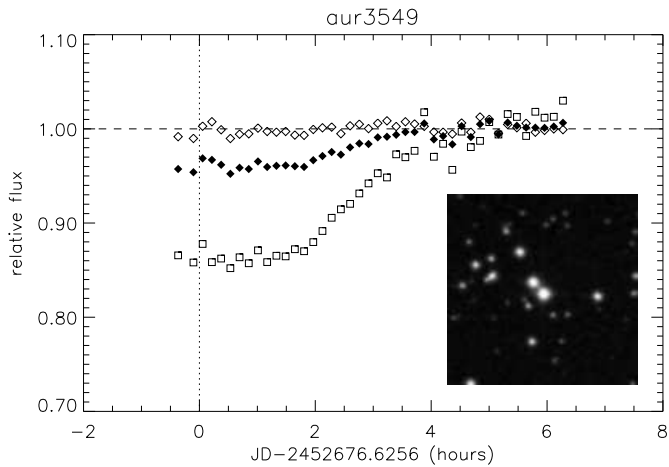}
\label{Looper}
\caption{The inset Digitized Sky Survey image shows this target to consist of 
a central bright source and an adjacent fainter star to the NE.  Both of 
these stars are contained within the PSF of the PSST. We carried out 
follow-up photometry of this field with a 14-inch telescope located
on the roof of the Caltech Astronomy building, during a night when an eclipse
was predicted to occur. Observations began during mid-eclipse. The predicted 
time of eclipse occurs at t=0 of this figure. The open diamonds show the 
light curve for the central star, which is evidently uneclipsed. The open 
squares show the light curve for the star to the NE, which undergoes a deep 
(14 \%) eclipse. When the light from both stars is summed in a single 
aperture, the apparent eclipse depth is reduced to $\sim 3$ \%, reproducing 
the PSST light curve (see Figure 2).}
\end{figure}

\section{Sherlock: Removing the Eclipsing Binaries}

We are currently assembling a telescope named Sherlock that will be dedicated 
to examining transit candidates from Sleuth, as well as a number of other 
transit surveys which monitor stars brighter than V=13. Highly automated and 
observing in custom RGB filters with a better angular resolution (1.7 
arcsec/pixel) than Sleuth (10 arcsec/pixel), Sherlock will be able to reject 
most of the contaminants from these transit surveys. In so doing, Sherlock 
will greatly reduce the rate of false positives, bringing the number of 
transit candidates to a manageable level. Figures 2 \& 3 show an example of 
how blends of eclipsing binaries can be separated from transiting planets. 
PSST identified a star in Auriga undergoing 2.5 \% deep eclipses with a 
period of 2.4 days (Figure 2). Examination of the Digitized Sky Survey images 
(Figure 3) revealed two sources within the PSF of the PSST instrument. 
Photometry of the field during a subsequent eclipse with a 14-inch telescope 
on the rooftop of the Caltech Astronomy building demonstrated that the 
fainter star was undergoing deep (14 \%) eclipses. When the light from both 
stars was summed within a single photometric aperture, the transit shape and 
depth as observed by the PSST was reproduced (Figure 3).

Measurement of the color dependence of the transit depth should remove both 
grazing incidence binaries (due to the effect of limb-darkening on the 
eclipse depth), and blends of eclipsing binaries (due the change in relative 
brightness of the blended and occulted stars as a function of color). In 
contrast to this, planetary transits should be nearly constant with color. 
Furthermore, the increased angular resolution will separate the light from 
physically unassociated blended stars, and subsequent photometry will reveal 
which object is undergoing eclipses. Sherlock will not be able to reject all 
sources of false detection. Central eclipses by very dim stellar objects 
(notably M dwarfs) will not show detectable variation with color. In 
addition, blends wherein the occulted and blending star have the same color 
will not be distinguished. Even with these exceptions, however, Sherlock is 
expected to reduce greatly the ratio of false positives from these surveys. 
Multi-epoch spectra of viable candidates will be gathered with 
high-resolution spectrographs on 1-2 m class telescopes, which should rule 
out the presence of stellar or brown dwarf companions. Surviving candidates 
will be monitored with Keck HIRES to determine the radial velocity orbit 
induced by the planetary companion.

\section{Sherlock Specifications}

\begin{table}
\begin{tabular}{| l |}
\hline
Meade LX200GPS 10'' f/6.3 Schmidt-Cassegrain Telescope \\
Apogee 1024 x 1024 pixel back-illuminated CCD camera \\
Filter wheel containing custom RGB and clear filters \\
SBIG STV Autoguider \\
Automated operation controlled by Linux workstation \\
Cloud cover monitored by Snoop, the Palomar All-Sky Camera \\
\hline
\end{tabular}
\caption{Specifications of Sherlock}
\label{specs}
\end{table}

Table~\ref{specs} lists the specifications for Sherlock. The system will be 
located in the same clamshell enclosure as Sleuth, our primary transit search 
instrument. Weather decisions are made by the on-site 200-inch telescope night
assistant, with additional protection provided by a weather station capable of
closing the clamshell roof. Sherlock will be completely automated, 
calculating future times of eclipse for all active candidates, and observing 
the highest priority object in eclipse each night. The automated nature of the
system is an advantage over comparatively labor- and resource-intensive 
multi-epoch spectroscopic follow-up. In addition, this dome hosts the new 
all-sky camera Snoop~\footnote{\url{http://snoop.palomar.caltech.edu}}, which 
provides weather monitoring for the observatory. Given that observing time on 
this system should be plentiful, we invite teams conducting wide-field 
transit surveys to contact us regarding follow-up of their candidates.

\bibliographystyle{aipproc}

\end{document}